\documentclass[final,3p,times,twocolumn]{elsarticle}

\usepackage{amsmath}
\usepackage{amssymb}
\usepackage[font=footnotesize]{caption}
\usepackage[colorlinks=true]{hyperref}
\usepackage[capitalize]{cleveref}
\usepackage[numbers]{natbib}
\usepackage[font=footnotesize]{subcaption}

\newcommand{\C}{\mathbb{C}}
\newcommand{\eps}{\varepsilon}
\newcommand{\ii}{\mathrm{i}}
\newcommand{\SU}{\mathrm{SU}}
\newcommand{\SL}{\mathrm{SL}}

\journal{Journal of Subatomic Particles and Cosmology}

\begin{document}

\begin{frontmatter}

\title{Hot and dense QCD equation of state from the complex Langevin equation}

\author[aaa]{Michael Mandl\corref{cor1}}
\cortext[cor1]{Corresponding author}
\ead{michael.mandl@uni-graz.at}

\author[aaa]{Dénes Sexty}
\ead{denes.sexty@uni-graz.at}

\author[aaa]{Daniel Unterhuber}
\ead{daniel.unterhuber@uni-graz.at}

\affiliation[aaa]{organization={Institute of Physics, NAWI Graz, University of Graz}, addressline=Universitätsplatz 5, city=8010, citysep={}, postcode=Graz, country=Austria}

\begin{abstract}
We discuss the QCD equation of state at temperatures above the crossover and for high baryon chemical potentials, obtained via the first ever complex Langevin simulations at the physical pion mass. In particular, we present continuum-extrapolated results for the energy, entropy, and enthalpy densities, as well as for the specific heat capacity as a function of baryon chemical potential and temperature.
\end{abstract}


\end{frontmatter}

\section{Introduction}The study of quantum chromodynamics (QCD) at non-zero baryon density using \emph{ab-initio} lattice simulations is impeded by a sign problem (see, e.g., \cite{AS26r}), preventing the application of conventional lattice techniques that rely on importance sampling. For this reason, non-perturbative knowledge about finite-density QCD, which is relevant for the theoretical description of, say, neutron stars or heavy-ion collisions, from first principles is severely limited. While approaches employing traditional hybrid-Monte-Carlo (HMC) simulations, which can study small non-zero baryon chemical potentials $\mu_B$ for instance via reweighting techniques, imaginary $\mu_B$, or Taylor expansions in $\mu_B/T$ (with $T$ denoting the temperature), have made substantial progress in recent years, they are limited by the ratio of chemical potential over temperature they can reliably reach. Currently, their range of applicability is restricted to $\mu_B/T\lesssim3.5$ (see, e.g., \cite{BGK22}).

In this work, we avoid the sign problem by employing the complex Langevin equation \cite{Par83,Kla83} to compute the equation of state of QCD at high temperatures and ratios of baryon chemical potential to temperature up to $\mu_B/T=10$. Building on our previous works \cite{MSU26,MSU26p}, where we have computed the baryon density, pressure, and trace anomaly, in this study we present results for the energy, entropy, and enthalpy densities, as well as the specific heat capacity, in the $(\mu_B,T)$ plane. We note, however, that our results are currently limited to temperatures well above the QCD crossover.

\section{Numerical setup}Our simulation setup is outlined in \cite{MSU26}. Here, we merely summarize the most important points. The complex Langevin approach, when applied to lattice QCD, relies on the complexification of the compact lattice gauge links $U_{x\mu}\in\SU(3)$ to non-compact $\SL(3,\C)$-valued ones and on the stochastic evolution of the latter in an auxiliary time dimension $\tau$, called the Langevin time. To be concrete, the $\tau$-evolution of the non-unitary gauge links can be simulated by employing the discrete update step
\begin{equation}\label{eq:cle}
	U_{x\mu}^{(\tau+\Delta\tau)} = \exp\left[\ii\sum_{a=1}^{8}\lambda_aM_{ax\mu}^{(\tau)}[U]\right]U_{x\mu}^{(\tau)}\;.
\end{equation}
Here, $U_{x\mu}^{(\tau)}$ denotes the gauge link $U_{x\mu}$, connecting lattice site $x$ with its nearest neighbor in $\mu$-direction, at Langevin time $\tau$, $\Delta\tau$ is a (small) discrete time-step, and $\lambda_a$ are the usual Gell-Mann matrices. Moreover, we have introduced the abbreviation
\begin{equation}\label{eq:exponent}
	M_{ax\mu}^{(\tau)}[U] = -\Delta\tau D_{ax\mu}S[U] + \sqrt{2\Delta\tau}\,\eta_{ax\mu}^{(\tau)}\;,
\end{equation}
where $S[U]$ is the (Euclidean) effective action of QCD once the fermions have been integrated out, $D_{ax\mu}$ denotes the usual left derivative, acting on functions $f$ of the gauge links as
\begin{equation}
	D_{x\mu a}f[U] = \frac{\partial}{\partial\alpha} f\left[e^{\ii\alpha\lambda_a}U_{x\mu}\right]\bigg\vert_{\alpha=0}\;,
\end{equation}
and $\eta_{ax\mu}^{(\tau)}$ are real and uncorrelated Gaussian noise variables with zero mean and unit variance. The update scheme \eqref{eq:cle} with \eqref{eq:exponent} is just one way of discretizing the continuum complex Langevin equation for lattice QCD, whereas for the present work we, in fact, employ an improved prescription \cite{UF85}, see \cite{MSU26}. Furthermore, we determine $\Delta\tau$ adaptively \cite{AJS10} and make use of the gauge-cooling technique \cite{SSS13} to stabilize the stochastic evolution. Since we work at high temperatures, our simulations do not exhibit excursions far into the non-unitary domain, implying that our results should be unaffected by the wrong-convergence problem that sometimes affects complex Langevin dynamics \cite{SSS20_2}.

We consider $2+1+1$ quark flavors and a physical pion mass, employing the staggered discretization for the fermions, a tree-level improved Symanzik action for the gauge fields, and four steps of stout smearing \cite{MP04,Sex19}. Our lattices have sizes $20^3\times10$, $24^3\times12$, and $32^3\times16$, which we use to perform a continuum extrapolation along the line of constant physics (LCP) computed in \cite{BBF15}, which employs an analogous setup (albeit on the back of more conventional HMC simulations). By performing cross-checks with larger volumes, we have confirmed that finite-size effects are under control in our simulations despite the relatively small aspect ratio of $N_s/N_t=2$, with $N_t$ and $N_s$ denoting the number of lattice points in the temporal and each spatial direction, respectively. Finally, we choose the quark chemical potentials in such a way to realize the strangeness and charm neutrality conditions as well as isospin symmetry, see \cite{MSU26,MSU26p}.

\section{Results}In our recent works \cite{MSU26,MSU26p}, we have presented results for the baryon density $n_B$, pressure $p$, and trace anomaly $I$ in the high-temperature region of the $(\mu_B,T)$ plane. Here, we use these quantities to compute additional thermodynamic observables relevant for the equation of state, namely the energy density $\eps$, entropy density $s$, enthalpy density $w$, and specific heat capacity $c_V$. Throughout, we use hats to denote quantities normalized by appropriate powers of the temperature to make them dimensionless, i.e.,
\begin{equation}
	\hat{\mu}_B = \frac{\mu_B}{T}\;, \ \ 
	\hat{\eps} = \frac{\eps}{T^4}\;, \ \ 
	\hat{s} = \frac{s}{T^3}\;, \ \ 
	\hat{w} = \frac{w}{T^4}\;, \ \ 
	\hat{c}_V = \frac{c_V}{T^3}\;.
\end{equation}
Moreover, we introduce the notation $\Delta X$ for the difference of a quantity $X$ with respect to its value at $\mu_B=0$ as 
\begin{equation}\label{eq:delta}
	\Delta X(T,\mu_B) = X(T,\mu_B) - X(T,0)\;.
\end{equation}
For the observables discussed in this work, we have direct access only to $\Delta X$. In order to compute the full quantities $X(T,\mu_B)$ from \eqref{eq:delta}, we take $X(T,0)$ as input from (the Supplementary Information of) \cite{BFG16}, either directly from the respective results (if available), or by performing simple computations analogous to those outlined below.

\begin{figure}[t]
	\centering
	\begin{subfigure}{\linewidth}
		\includegraphics[scale=0.48]{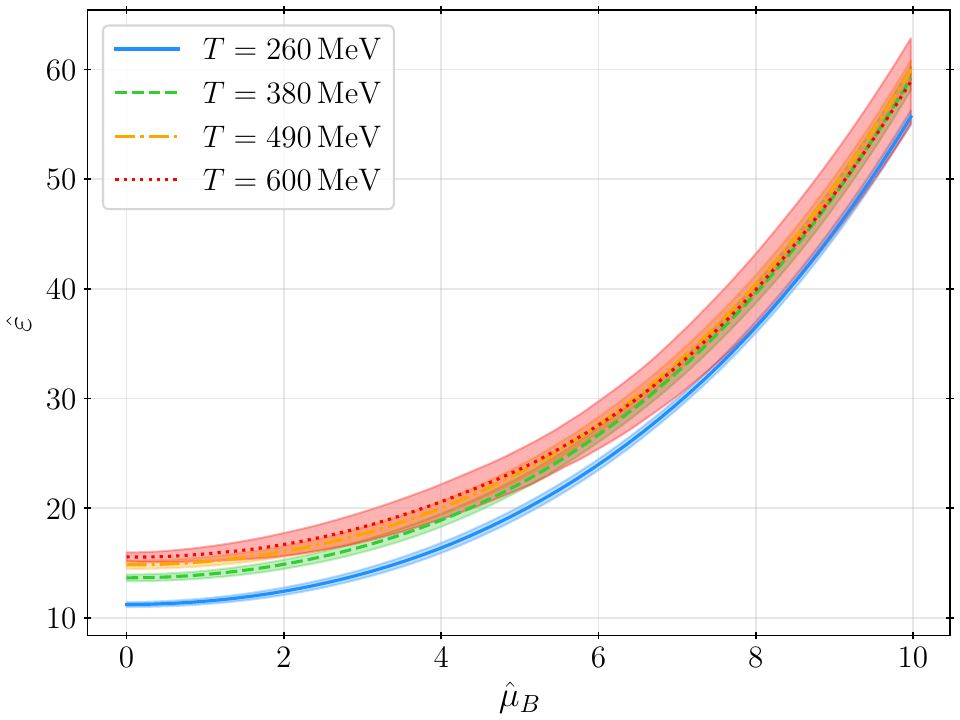}
	\end{subfigure}
	\begin{subfigure}{\linewidth}
		\includegraphics[scale=0.48]{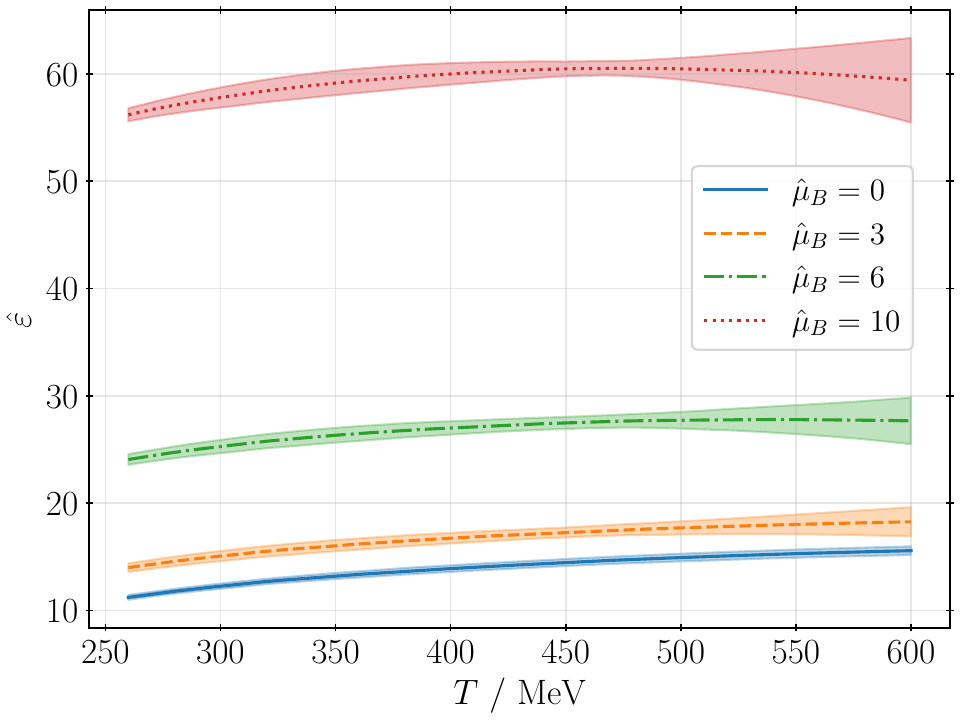}
	\end{subfigure}
	\caption{Dependence of the dimensionless energy density $\hat{\eps}$ on $\hat{\mu}_B$ for different $T$ \emph{(top)} and its dependence on $T$ for different $\hat{\mu}_B$ \emph{(bottom)}.}
	\label{fig:e}
\end{figure}

We begin by computing the energy density, defined as
\begin{equation}
	\eps = I + 3p\;.
\end{equation}
Results for the normalized energy density $\hat{\eps}$ as a function of $\hat{\mu}_B$ for different fixed temperatures, and as a function of $T$ for different fixed $\hat{\mu}_B$ are shown in \cref{fig:e}. One observes an increase with $\mu_B$ for all temperatures as well as an increase with $T$, at least for small chemical potentials. By comparing with \cite{MSU26p}, it becomes clear that the contribution of the pressure to the energy density dominates over the trace anomaly. For our highest temperature, i.e., $T=600\,\mathrm{MeV}$, the (combined statistical and systematic, see \cite{MSU26}) uncertainties become sizable. This is also true for the other observables studied in this work and can be attributed to systematic uncertainties arising in our fitting procedure for the temperature-dependence. 

\begin{figure}[t]
	\centering
	\begin{subfigure}{\linewidth}
		\includegraphics[scale=0.48]{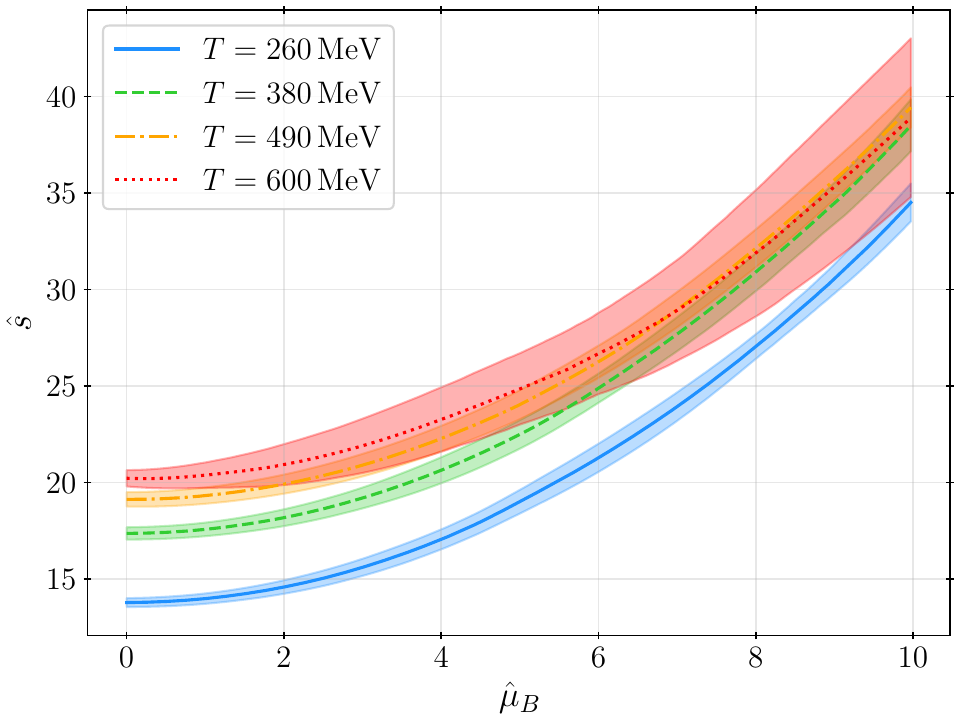}
	\end{subfigure}
	\begin{subfigure}{\linewidth}
		\includegraphics[scale=0.48]{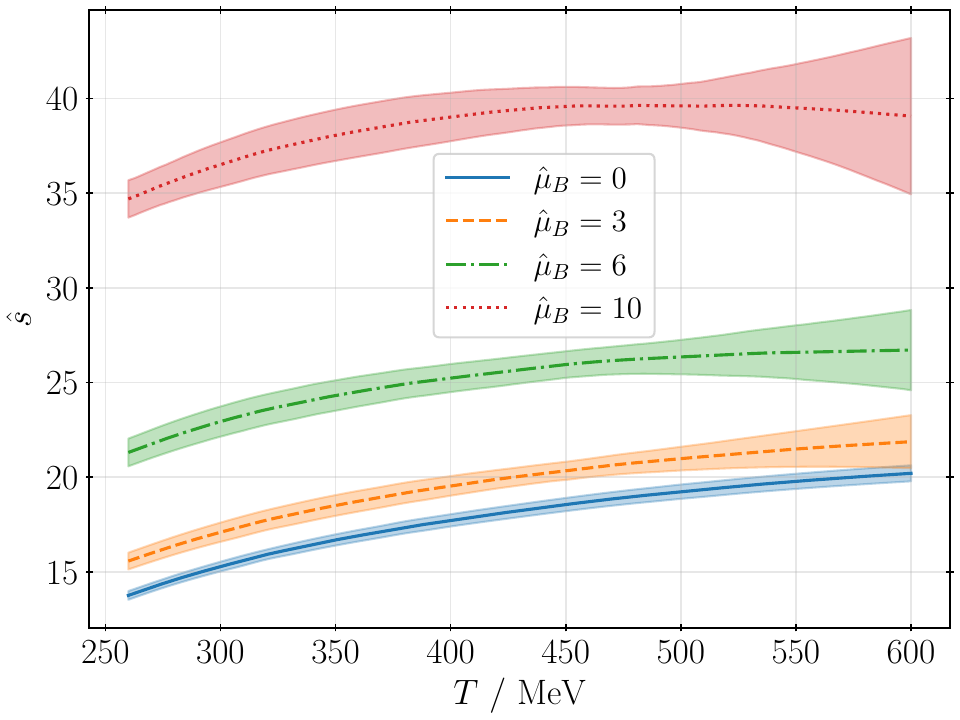}
	\end{subfigure}
	\caption{Same as \cref{fig:e} but for the entropy density $\hat{s}$.}
	\label{fig:s}
\end{figure}

\begin{figure}[t]
	\centering
	\begin{subfigure}{\linewidth}
		\includegraphics[scale=0.48]{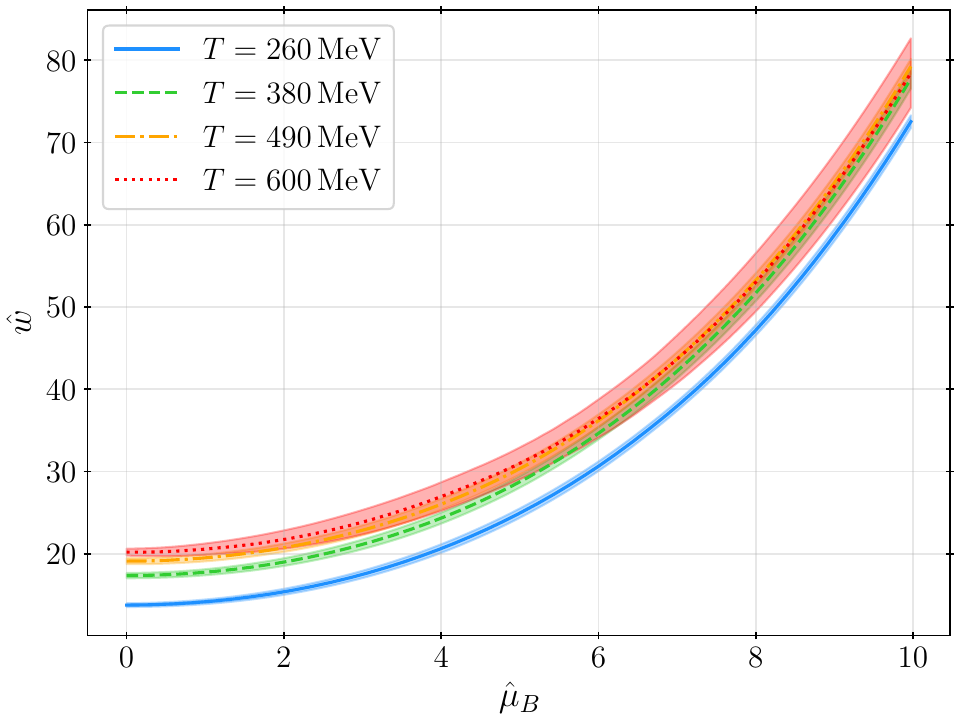}
	\end{subfigure}
	\begin{subfigure}{\linewidth}
		\includegraphics[scale=0.48]{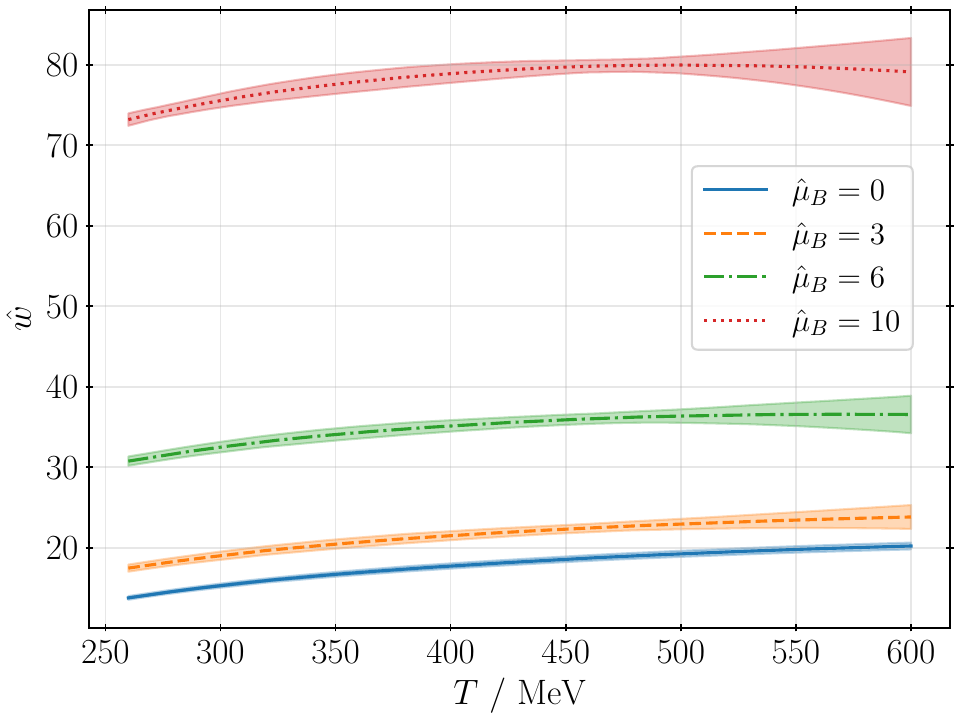}
	\end{subfigure}
	\caption{Same as \cref{fig:e} but for the enthalpy density $\hat{w}$.}
	\label{fig:w}
\end{figure}

Next, we discuss the entropy density
\begin{equation}
	s = \frac{1}{T}\left(\eps + p - \mu_B n_B\right)
\end{equation}
by showing $\hat{s}$ as a function of both $\hat{\mu}_B$ and $T$ in \cref{fig:s}. The qualitative behavior is very similar to that of the energy density in \cref{fig:e}. While at higher temperatures the data also allow for a change of monotonicity within the uncertainties, such a scenario seems to be implausible on physical grounds. The same conclusions are valid also for the enthalpy density
\begin{equation}
	w = \eps + p\;,
\end{equation}
whose normalized form is shown as a function of $\hat{\mu}_B$ and $T$ in \cref{fig:w}.

\begin{figure}[t]
	\centering
	\begin{subfigure}{\linewidth}
		\includegraphics[scale=0.48]{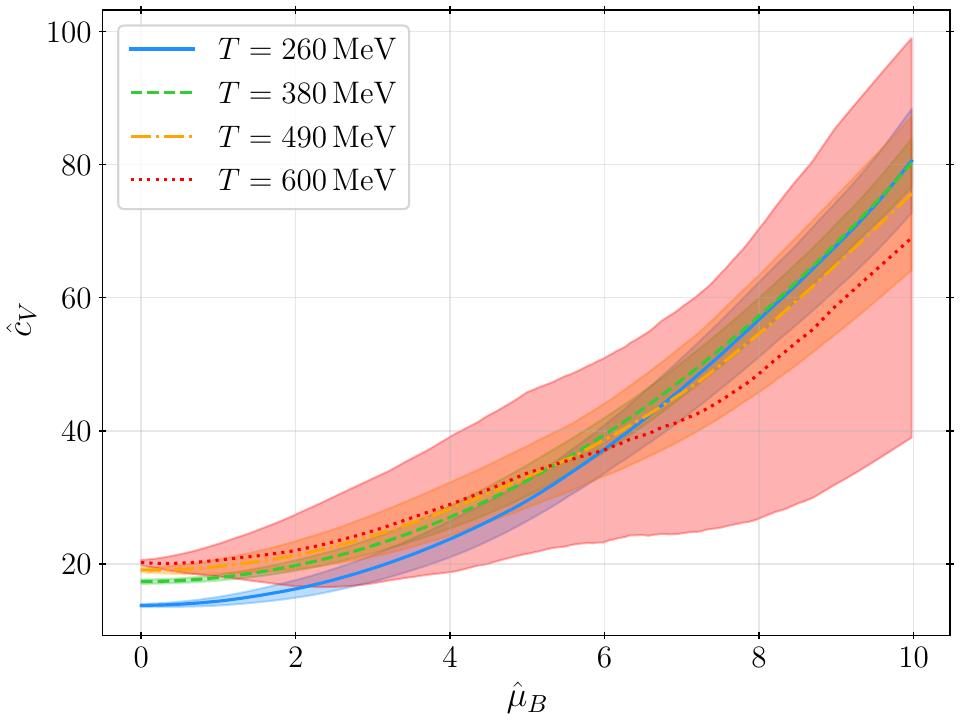}
	\end{subfigure}
	\begin{subfigure}{\linewidth}
		\includegraphics[scale=0.48]{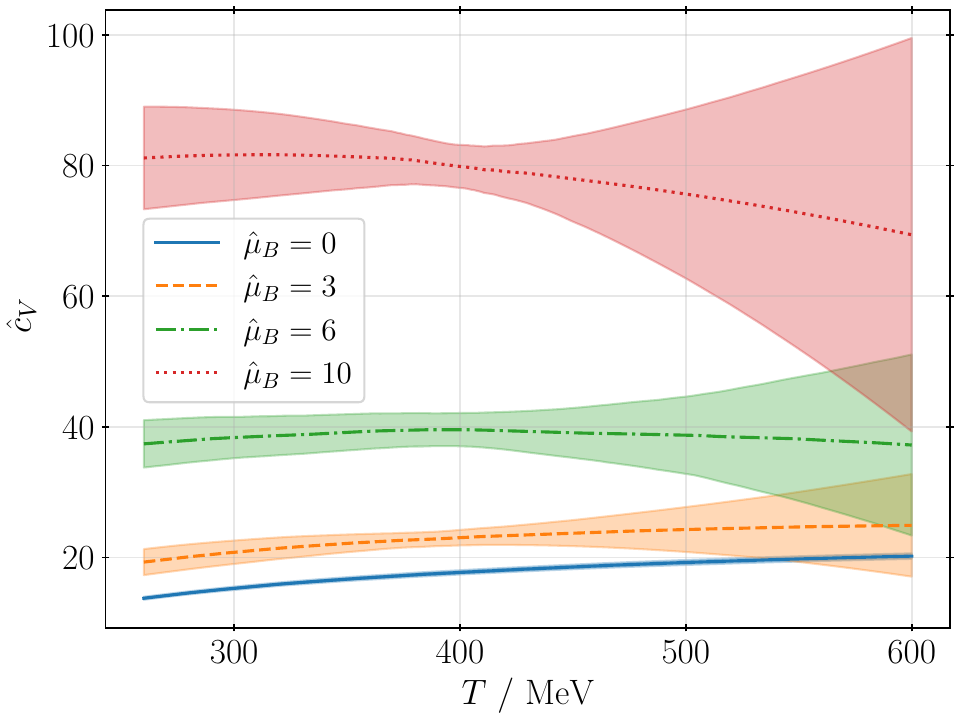}
	\end{subfigure}
	\caption{Same as \cref{fig:e} but for the specific heat capacity $\hat{c}_V$.}
	\label{fig:c}
\end{figure}

Finally, we consider the specific heat capacity,
\begin{equation}\label{eq:c}
	c_V = \frac{\partial\eps}{\partial T}\;,
\end{equation}
where the $T$-derivative is understood to be taken at constant spatial volume. In our calculations, however, we do not enforce this condition. Nonetheless, we expect the ensuing error to be negligible since we consider intensive quantities exclusively and finite-size effects are close to zero. In terms of the normalized specific heat capacity, \eqref{eq:c} can be written as
\begin{equation}
	\hat{c}_V = 4\hat{\eps} + T\frac{\partial\hat{\eps}}{\partial T} - \hat{\mu}_B\frac{\partial\hat{\eps}}{\partial\hat{\mu}_B}\;,
\end{equation}
which we evaluate numerically after fitting $\hat{\eps}$ in both $T$ and $\hat{\mu}_B$. The normalized specific heat capacity as a function of both $\hat{\mu}_B$ and $T$ is shown in \cref{fig:c}. The data suggests an increase of $\hat{c}_V$ with $\mu_B$ for low temperatures. For higher temperatures, on the other hand, the uncertainties become very large and it becomes difficult to determine the monotonicity adequately.

\section{Summary \& Outlook}We have calculated the energy density, entropy density, enthalpy density, and specific heat capacity in strangeness- and charm-neutral and isospin-symmetric $(2+1+1)$-flavor QCD for temperatures above the crossover and high baryon densities, reaching $\mu_B$-to-$T$ ratios up to $\mu_B/T=10$, therefore vastly extending the range of chemical potentials that is currently achievable with HMC-based lattice approaches. For our continuum-extrapolated lattice results, we have employed the first ever complex Langevin simulations at the physical point.

\section*{Acknowledgments}
We extend our gratitude to Astrid Carstensen, Erhard Seiler, and Ion-Olimpiu Stamatescu for regular discussions and collaborations past and present. Moreover, we thank Szabolcs Borsányi and Jana Günther for sharing the data for vanishing chemical potential and for the LCP with us. M.M.\ is grateful for enlightening discussions with Gert Aarts and Jana Günther. This research was funded in whole or in part by the Austrian Science Fund (FWF) \href{https://doi.org/10.55776/P36875}{10.55776/P36875}. The computational results have been obtained via the Austrian Scientific Computing (ASC) infrastructure and the computing cluster of the University of Graz (GSC).

\bibliographystyle{elsarticle-num}
\bibliography{bibliography}

\end{document}